\documentclass[sort&compress
    ,final            % use final for the camera ready runs
  ]
  {aipproc}

\layoutstyle{6x9}

\usepackage{amsmath}
\usepackage{amssymb}
\newcommand{\SU}{\mathop{\rm SU}\nolimits}

\newcommand{\eq}[1]{eq.~(\ref{#1})}
\newcommand{\Eq}[1]{eq.~(\ref{#1})}
\begin{document}

\title{Finite $SU(3)^3$ model}


\author{S.~Heinemeyer}{
  address={Instituto de Físicade Cantabria (CSIC-UC), Santander, Spain }
}

\author{E.~Ma}{
 address={Physics Department, University of California, 
	Riverside, California 92521, U.S.A.}
}

\author{M.~Mondragon}{
  address={Inst. de F\'{\i}sica, 
    Universidad Nacional Aut\'onoma de M\'exico (IF-UNAM), M\'exico D.F.}
 % additional visiting address
}

\author{G.~Zoupanos}{
  address={Physics Department, National Technical University,
GR-157 80 Zografou, Athens, Greece.}
 % additional visiting address
}

\begin{abstract}
  We consider N=1 supersymmetric gauge theories based on the group
  $SU(N)_1 \times SU(N)_2 \times ... \times SU(N)_k$ with matter
  content $(N,N^*,1,...,1) + (1,N,N^*,...,1) + ... + (N^*,1,1,...,N)$
  as candidates for the unification symmetry of all particles. In
  particular we examine to which extent such theories can become
  finite, and find that a necessary condition is that there should be
  exactly three families.  From phenomenological considerations an
  $SU(3)^3$ model is singled out. We consider an all-loop and a
  two-loop finite model based on this gauge group and we study their
  predictions concerning the third generation quark masses.
\end{abstract}

\maketitle

\section{Introduction}

Finite Unified Theories (FUTs) are $N=1$ supersymmetric Grand Unified
Theories (GUTs) which can be made finite to all-loop orders, including
the soft supersymmetry breaking sector.  The constructed {\it finite
  unified} $N=1$ supersymmetric SU(5) GUTs predicted correctly from
the dimensionless sector (Gauge-Yukawa unification), among others, the
top quark mass \cite{Kapetanakis:1992vx,Mondragon:1993tw}.
  Eventually, the full theories can be made all-loop finite and their
  predictive power is extended to the Higgs sector and the s-spectrum
  \cite{Heinemeyer:2007tz}. For a detailed discussion see
  \cite{Kubo:1997fi,Kobayashi:2001me}.

Consider  a chiral, anomaly free,
$N=1$ globally supersymmetric
gauge theory based on a group G with gauge coupling
constant $g$. The
superpotential of the theory is given by
\begin{equation}
 W= \frac{1}{2}\,m^{ij} \,\Phi_{i}\,\Phi_{j}+
\frac{1}{6}\,C^{ijk} \,\Phi_{i}\,\Phi_{j}\,\Phi_{k}~, 
\label{1}
\end{equation}
where $m^{ij}$ (the mass terms) and $C^{ijk}$ (the Yukawa couplings)
are gauge invariant tensors, and the matter field $\Phi_{i}$
transforms according to the irreducible representation $R_{i}$ of the
gauge group $G$.  All the one-loop $\beta$-functions of the theory
vanish if the $\beta$-function of the gauge coupling $\beta_g^{(1)}$,
and the anomalous dimensions of the Yukawa couplings
$\gamma_i^{j(1)}$, vanish, i.e.
\begin{equation}
\sum _i \ell (R_i) = 3 C_2(G) \,,~
\frac{1}{2}C_{ipq} C^{jpq} = 2\delta _i^j g^2  C_2(R_i)\ ,
\label{2}
\end{equation}
where $\ell (R_i)$ is the Dynkin index of $R_i$, and $C_2(G)$ is the
quadratic Casimir invariant of the adjoint representation of $G$. A
theorem given in
\cite{Lucchesi:1987he,Piguet:1986td,Piguet:1986pk,Lucchesi:1996ir}
then guarantees the vanishing of the $\beta$-functions to all-orders
in perturbation theory.  This requires that, in addition to the
one-loop finiteness conditions (\ref{2}), the Yukawa couplings are
reduced in favour of the gauge coupling.

In the soft breaking sector, it was found that the soft supersymmetry
breaking (SSB) scalar masses in Gauge-Yukawa and finite unified models
satisfy a sum rule \cite{Kobayashi:1998jq,Kubo:1994bj}
\begin{equation}
\frac{(~m_{i}^{2}+m_{j}^{2}+m_{k}^{2}~)}{M M^{\dag}} =
1+\frac{g^2}{16 \pi^2}\,\Delta^{(1)}
+O(g^4)~
\label{sumr}
\end{equation}
for i, j, k, where $\Delta^{(1)}$ is
the two-loop correction, which vanishes when all the soft scalar
masses are the same at the unification point.

\section{Finite $SU(3)^3$ model}
We now examine the possibility of constructing realistic FUTs
based on product gauge groups. Consider an $N=1$ supersymmetric
theory, with gauge group $\SU(N)_1
\times \SU(N)_2 \times \cdots \times \SU(N)_k$, with $n_f$ copies of
the supersymmetric multiplet $(N,N^*,1,\dots,1) + (1,N,N^*,\dots,1) +
\cdots + (N^*,1,1,\dots,N)$.  The one-loop $\beta$-function
coefficient in the renormalization-group equation of each $\SU(N)$
gauge coupling is simply given by
\begin{equation}
b = \left( -{11 \over 3} + {2 \over 3} \right) N + n_f \left( {2 \over
  3} + {1 \over 3} \right) \left( {1 \over 2} \right) 2 N = -3 N + n_f
N\,.
\label{3gen}
\end{equation}
This means that $n_f = 3$ is the only solution of \Eq {3gen} that
yields $b = 0$.  Since $b=0$ is a
necessary condition for a finite field theory, the existence of three
families of quarks and leptons is natural in such models, provided the
matter content is exactly as given above.

The model of this type with best phenomenology is the $SU(3)^3$ model
discussed in ref.~\cite{Ma:2004mi}, where the details of the model are
given. It corresponds to the well-known example of $\SU(3)_C
\times \SU(3)_L \times
\SU(3)_R$~\cite{Derujula:1984gu,Lazarides:1993sn,Lazarides:1993uw,Ma:1986we}.
%\section{An all-loop $\SU(3)^3$ FUT}\label{section5}
Thus, this $\SU(3)^3$ model is finite between the Planck $M_P$ and the
unification $M_{GUT}$ scales, then breaks spontaneously down to the
MSSM at $M_{GUT}$ \cite{Lazarides:1993uw}. Notice that this model has
extra exotic particles above $M_{GUT}$, which are Higgs-like and
down-quark like.

With three families, the most general superpotential contains 11 $f$
couplings, and 10 $f'$ couplings, subject to 9 conditions, due to the
vanishing of the anomalous dimensions of each superfield.  The
conditions are the following
\begin{equation}
\sum_{j,k} f_{ijk} (f_{ljk})^* + \frac{2}{3} \sum_{j,k} f'_{ijk}
(f'_{ljk})^* = \frac{16}{9} g^2 \delta_{il}\,,
\label{19}
\end{equation}
We will assume that the below $M_{GUT}$, we have the usual MSSM, with
the two Higgs doublets coupled maximally to the third generation.
The remnants of the $\SU(3)^3$ FUT are the boundary conditions on the
gauge and Yukawa couplings, i.e. \eq{19}, the $h=-MC$
relation, and the soft scalar-mass sum rule~(\ref{sumr}) at $M_{\rm
  GUT}$, which, when applied to the present model, takes the form
\begin{eqnarray}
m^2_{H_u} + m^2_{\tilde t^c} + m^2_{\tilde q} = M^2 = 
m^2_{H_d} + m^2_{\tilde b^c} + m^2_{\tilde q}.
\end{eqnarray}

Concerning the solution to \eq{19} we consider two versions of the model:\\
I) An all-loop finite model with a unique and isolated solution, in
which $f'$ vanishes, which leads to the following relation
\begin{equation}
f^2 = f^2_{111} = f^2_{222} = f^2_{333} = \frac{16}{9} g^2\,.
\label{isosol}
\end{equation}
  As for
the lepton masses, because all $f'$ couplings have been fixed to be
zero at this order, in principle they would be expected to appear
radiatively induced by the scalar lepton masses appearing in the SSB
sector of the theory.  However, due to the finiteness
conditions they cannot appear radiatively and remain as a
problem for further study.  \\ 
II) A two-loop finite solution, in which we keep $f'$
  non-vanishing and we use it to introduce the lepton
  masses. The model in turn becomes finite
only up to two-loops since the corresponding solution of \Eq{19} is
not isolated. In this case we have the following
boundary conditions for the Yukawa couplings 
\begin{eqnarray}
f^2 = r \left(\frac{16}{9}\right) g^2\,,\quad  
f'^2 = (1-r) \left(\frac{8}{3}\right) g^2\,.
\label{fprime}
\end{eqnarray}
As for the boundary conditions of the soft scalars, we have the
universal case.

\section{Predictions and conclusions}\label{section6}

\begin{figure}
\includegraphics[width=6cm,angle=0]{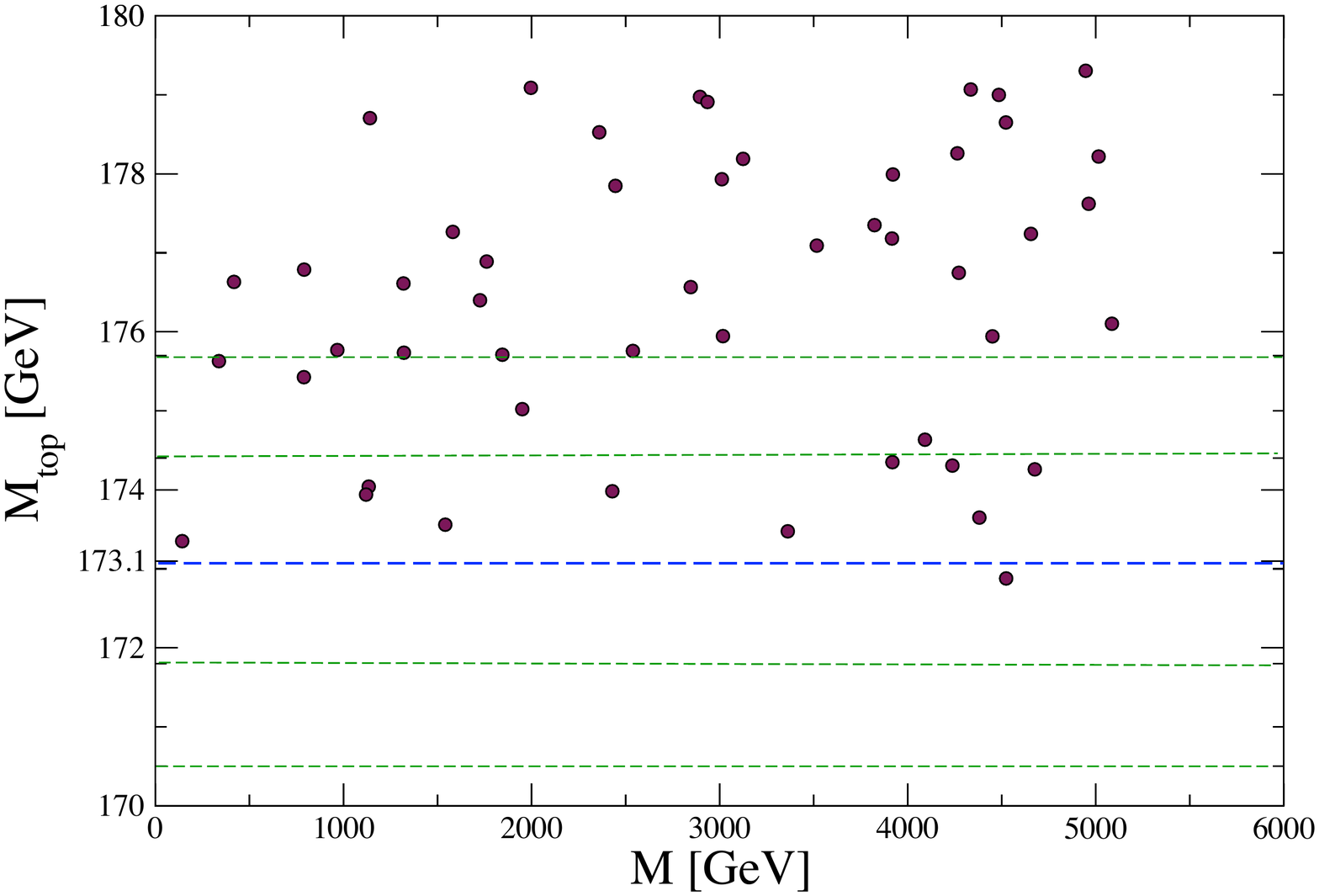}
\includegraphics[width=6cm,angle=0]{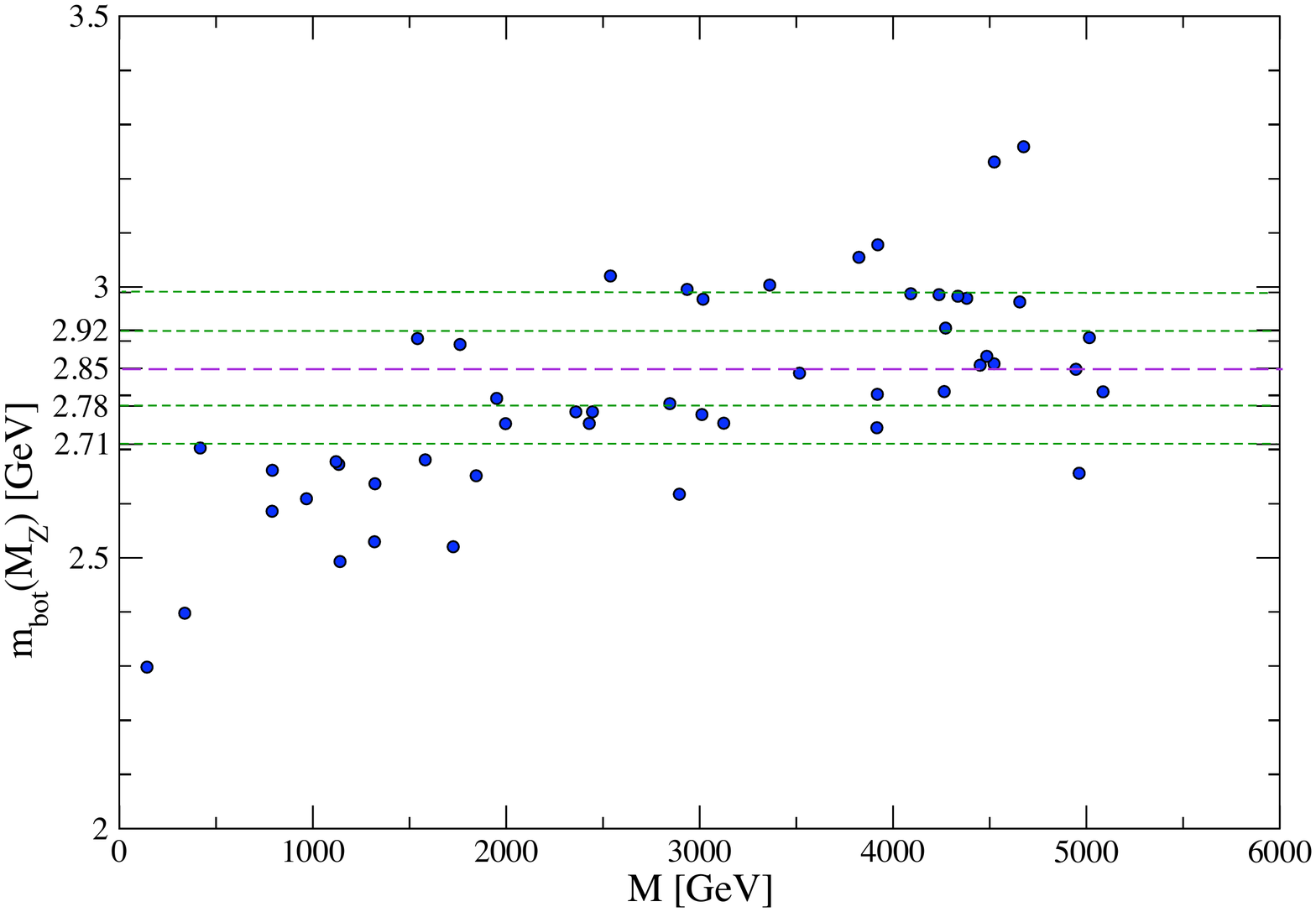}
\caption{The figure shows the values for the top and bottom quark
  masses for model II, with $r \sim 0.7$ and $\mu <0 $.}
\label{fig:Higgs-m5}
\end{figure}

Below $M_{GUT}$ all couplings and masses of the theory run according
to the RGEs of the MSSM.  Thus we examine the evolution of these
parameters according to their RGEs up to two-loops for dimensionless
parameters and at one-loop for dimensionful ones imposing the
corresponding boundary conditions.  We further assume a unique
supersymmetry breaking scale $M_{s}$ and below that scale the
effective theory is just the SM.

We compare our predictions with the most recent experimental value $
m_t^{exp} = (173.1 \pm 1.3)\,{\rm ~GeV}$~\cite{:2009ec}, and recall
that the theoretical values for $m_t$ suffer from a correction of
$\sim 4 \%$~\cite{Kubo:1997fi,Kobayashi:2001me}. In the case of the
bottom quark, we take the value evaluated at $M_Z$, $m_b
(M_Z)=2.85\pm0.07\,{\rm GeV}$~\cite{Amsler:2008zzb}.  In the case of
model I, the predictions for the top quark mass (in this case $m_b$ is
fixed) $m_t$ are $\sim 183\,{\rm ~GeV}$ for $\mu < 0 $, which is above
the experimental value.

For the two-loop model {II}, we look for the values of the parameter
$r$ which comply with the experimental limits given above for top and
bottom quarks masses. In the case of $\mu >0$, for the bottom quark,
the values of $r$ lie in the range $0.15 \lesssim r \lesssim 0.32$.
For the top mass, the range of values for r is $0.35 \lesssim r \lesssim
0.6$. From these values we can see that there is a very small region
where both top and bottom quark masses are in the experimental range for
the same value of $r$.
In the case of $\mu<0$ the situation is similar, although slightly
better, with the range of values $0.62 \lesssim r \lesssim 0.77$ for
the bottom mass, and $0.4 \lesssim r \lesssim 0.62$ for the top quark
mass. In this case, if we take some of the exotic particles into
account, decoupling a bit below the unification scale, the situation
improves.  This can be seen in Fig. 1, where we took three down-like and
one Higgs-like exotic particles between $10^{15}$ and $10^{16}$ GeV,
below that the usual MSSM. Then, for $r\sim 0.7$ we have good
agreement with experimental data for both top and bottom quarks
\cite{new-HMMZ}.

\vspace{1cm}
%\section*{Acknowledgements}
This work is partially supported by the NTUA's basic research support 
programme 2008 and the European Union's RTN programme under contract 
MRTN-CT-2006-035505.
Supported also by a mexican PAPIIT grant IN111609.


\begin{thebibliography}{19}
\expandafter\ifx\csname natexlab\endcsname\relax\def\natexlab#1{#1}\fi
\providecommand{\enquote}[1]{``#1''}
\expandafter\ifx\csname url\endcsname\relax
  \def\url#1{\texttt{#1}}\fi
\expandafter\ifx\csname urlprefix\endcsname\relax\def\urlprefix{URL }\fi
\providecommand{\eprint}[2][]{\url{#2}}

\bibitem[Kapetanakis et~al.(1993)]{Kapetanakis:1992vx}
D.~Kapetanakis, M.~Mondragon, and G.~Zoupanos, \emph{Z. Phys.} \textbf{C60},
  181--186 (1993), \eprint{hep-ph/9210218}.

\bibitem[Mondragon and Zoupanos(1995)]{Mondragon:1993tw}
M.~Mondragon, and G.~Zoupanos, \emph{Nucl. Phys. Proc. Suppl.} \textbf{37C},
  98--105 (1995).

\bibitem[Heinemeyer et~al.(2008)]{Heinemeyer:2007tz}
S.~Heinemeyer, M.~Mondragon, and G.~Zoupanos, \emph{JHEP} \textbf{07}, 135
  (2008), \eprint{0712.3630}.

\bibitem[Kubo et~al.(1997)]{Kubo:1997fi}
J.~Kubo, M.~Mondragon, and G.~Zoupanos, \emph{Acta Phys. Polon.} \textbf{B27},
  3911--3944 (1997), \eprint{hep-ph/9703289}.

\bibitem[Kobayashi et~al.(2001)]{Kobayashi:2001me}
T.~Kobayashi, J.~Kubo, M.~Mondragon, and G.~Zoupanos, \emph{Surveys High Energ.
  Phys.} \textbf{16}, 87--129 (2001).

\bibitem[Lucchesi et~al.(1988)]{Lucchesi:1987he}
C.~Lucchesi, O.~Piguet, and K.~Sibold, \emph{Helv. Phys. Acta} \textbf{61}, 321
  (1988).

\bibitem[Piguet and Sibold(1986{\natexlab{a}})]{Piguet:1986td}
O.~Piguet, and K.~Sibold, \emph{Int. J. Mod. Phys.} \textbf{A1}, 913
  (1986{\natexlab{a}}).

\bibitem[Piguet and Sibold(1986{\natexlab{b}})]{Piguet:1986pk}
O.~Piguet, and K.~Sibold, \emph{Phys. Lett.} \textbf{B177}, 373
  (1986{\natexlab{b}}).

\bibitem[Lucchesi and Zoupanos(1997)]{Lucchesi:1996ir}
C.~Lucchesi, and G.~Zoupanos, \emph{Fortschr. Phys.} \textbf{45}, 129--143
  (1997), \eprint{hep-ph/9604216}.

\bibitem[Kobayashi et~al.(1998)]{Kobayashi:1998jq}
T.~Kobayashi, J.~Kubo, and G.~Zoupanos, \emph{Phys. Lett.} \textbf{B427},
  291--299 (1998), \eprint{hep-ph/9802267}.

\bibitem[Kubo et~al.(1994)]{Kubo:1994bj}
J.~Kubo, M.~Mondragon, and G.~Zoupanos, \emph{Nucl. Phys.} \textbf{B424},
  291--307 (1994).

\bibitem[Ma et~al.(2004)]{Ma:2004mi}
E.~Ma, M.~Mondragon, and G.~Zoupanos, \emph{JHEP} \textbf{12}, 026 (2004),
  \eprint{hep-ph/0407236}.

\bibitem[De~R{\'u}jula and L.(1984)]{Derujula:1984gu}
G.~A. De~R{\'u}jula, and G.~S. L. p.~88 (1984), fifth Workshop on Grand
  Unification, K.~Kang, H.~Fried, and P.~Frampton eds., World Scientific,
  Singapore.

\bibitem[Lazarides et~al.(1993)]{Lazarides:1993sn}
G.~Lazarides, C.~Panagiotakopoulos, and Q.~Shafi, \emph{Phys. Lett.}
  \textbf{B315}, 325--330 (1993), \eprint{hep-ph/9306332}.

\bibitem[Lazarides and Panagiotakopoulos(1994)]{Lazarides:1993uw}
G.~Lazarides, and C.~Panagiotakopoulos, \emph{Phys. Lett.} \textbf{B336},
  190--193 (1994), \eprint{hep-ph/9403317}.

\bibitem[Ma(1987)]{Ma:1986we}
E.~Ma, \emph{Phys. Rev.} \textbf{D36}, 274 (1987).

\bibitem[Group(2009)]{:2009ec}
T.~E.~W. Group  (2009), \eprint{0903.2503}.

\bibitem[Amsler et~al.(2008)]{Amsler:2008zzb}
C.~Amsler, et~al., \emph{Phys. Lett.} \textbf{B667}, 1 (2008).

\bibitem[Heinemeyer~S. and Zoupanos(????)]{new-HMMZ}
M.~Mondragon, S.~Heinemeyer, E.~Ma, and G.~Zoupanos, {to appear.}

\end{thebibliography}
\end{document}